# Topological edge solitons in $\chi^{(2)}$ waveguide arrays

**YAROSLAV V. KARTASHOV**[1,*]

[1]*Institute of Spectroscopy, Russian Academy of Sciences, 108840, Troitsk, Moscow, Russia*
*Corresponding author: Yaroslav.Kartashov@icfo.eu*

**We address the formation of $\chi^{(2)}$ topological edge solitons emerging in topologically nontrivial phase in Su-Schrieffer-Heeger (SSH) waveguide arrays. We consider edge solitons, whose fundamental frequency (FF) component belongs to the topological gap, while phase mismatch determines whether second harmonic (SH) component falls into topological or trivial forbidden gaps of the spectrum for SH wave. Two representative types of edge solitons are found, one of which is thresholdless and bifurcates from topological edge state in FF component, while other exists above power threshold and emanates from topological edge state in SH wave. Both types of solitons can be stable. Their stability, localization degree, and internal structure strongly depend on phase mismatch between FF and SH waves. Our results open new prospects for control of topologically nontrivial states by parametric wave interactions.**

Excitations in topological insulators possess distinctly different propagation properties in comparison with conventional insulators. Despite that topological systems may be insulating in the bulk, they support localized in-gap states at their edges that are protected by the topology and are remarkably robust to disorder. Topological insulators are observed in electronic systems, in mechanics, acoustics, and in photonic systems [1,2]. Among the advantages of photonic systems is that they can possess strong nonlinearity and this allows to investigate the whole spectrum of new phenomena appearing in topologically nontrivial systems due to light self-action or parametric wave interactions, see recent review [3].

Among such phenomena is the possibility of tuning frequencies and localization of topological edge states by nonlinearity [4], active tuning of power transfer in topological chains [5], development of instabilities of periodic edge states [6,7] indicating on the possibility of formation of travelling localized edge solitons [8], formation of unidirectional solitons in the bulk [9] and at the edge [7,10] of topological system that were observed in Floquet insulators based on waveguide arrays [11,12] (theory of such solitons is developed in [13] for discrete and in [14,15] for continuous systems), formation of topological gap solitons in Su-Schrieffer-Heeger arrays [16-19] (such structures, including trimer chains, can be fabricated also using split-ring resonators [20,21]), realization of nonlinearity-controlled edge state switching [22], unusual types of topological Dirac [23], Bragg [24], and valley-Hall [25,26] solitons, induction of topological phases by the nonlinearity [27], and realization of nonlinear second-order topological insulators [28,29].

While topological solitons in materials with Kerr-type nonlinearity were reported, less attention was paid to the formation of such states due to parametric interactions in $\chi^{(2)}$ materials. Interactions of the extended edge states in such media were reported in [30], but the only example of wide unidirectional edge solitons was presented in $\chi^{(2)}$ Floquet system [31].

In this Letter we report on edge solitons in topological dimerized SSH waveguide arrays with $\chi^{(2)}$ nonlinearity, where transition from trivial to topological phase is controlled by shift of the waveguides in each dimer forming the array. The peculiarity of this system is that FF and SH waves have distinct linear spectra, and even if FF component belongs to the topological gap, the SH component may fall into trivial or topological gap depending on the phase mismatch between waves that, in turn, determines internal structure of soliton. Two types of edge solitons were found, one of which is thresholdless and bifurcates from topological mode in FF component, while soliton emanating from topological SH mode features power threshold. These properties substantially differ from those of usual surface [32-34] solitons in $\chi^{(2)}$ waveguide arrays.

We describe propagation of light beams in SSH array created in $\chi^{(2)}$ material using parametrically coupled nonlinear equations for the dimensionless amplitudes of the FF, $\psi_1$, and SH, $\psi_2$, waves:

$$i\frac{\partial \psi_1}{\partial z} = -\frac{1}{2}\frac{\partial^2 \psi_1}{\partial x^2} - \psi_1^* \psi_2 - \mathcal{R}(x)\psi_1,$$
$$i\frac{\partial \psi_2}{\partial z} = -\frac{1}{4}\frac{\partial^2 \psi_2}{\partial x^2} - \psi_1^2 + \beta\psi_2 - 2\mathcal{R}(x)\psi_2. \qquad (1)$$

Here all quantities are dimensionless, the transverse coordinate $x$ is normalized to the characteristic scale $x_0$, propagation distance $z$ is scaled to the diffraction length $k_1 x_0^2$, $k_1 = n_1(\omega)\omega/c$ and $k_2 = n_2(2\omega)2\omega/c$ are the wavenumbers of the FF and SH waves, $\psi_1 = 2\pi\omega^2\chi^{(2)}x_0^2 c^{-2}E_1$ and $\psi_2 e^{i\beta z} = 2\pi\omega^2\chi^{(2)}x_0^2 c^{-2}E_2$, where $E_{1,2}$ are the dimensional fields, $\chi^{(2)}$ is the second-order susceptibility for the employed phase matching scheme, $\beta = (2k_1 - k_2)k_1 x_0^2$ is the normalized phase mismatch. SSH array $\mathcal{R}(x) = p\sum_{m=1,2n} e^{-[x+(n-m+1/2)a+(-1)^{m-1}s]^2/d^2}$ consists of $n$ dimers (in our case $n=60$) with two Gaussian waveguides of width $d$ in each dimer that can be shifted by the distance $s$ in the opposite directions. At $s=0$ the array is uniform with equal spacing $a$ between all guides. Here $p \sim \max(\delta\chi^{(1)})2\pi(\omega x_0/c)^2$ is the depth of the optical potential $\mathcal{R}$. The potential is two times stronger for the SH wave due to difference of carrier frequencies. Further we set $a=2$, $d=0.5$, $p=2.5$ that is consistent with experimental parameters for arrays created by Ti -indiffusion in LiNbO$_3$ crystals with low losses $\sim 0.2$ dB/cm [32]. Omitting nonlinear terms in Eqs. (1) we calculate linear spectra and modes for FF $[\psi_1=w_1(x)e^{ib_{FF}z}]$ and SH $[\psi_2=w_2(x)e^{ib_{SH}z}]$ waves from respective linear eigenproblems $b_{FF}w_1=(1/2)d^2w_1/dx^2+\mathcal{R}w_1$ and $b_{SH}w_2=(1/4)d^2w_2/dx^2-\beta w_2+2\mathcal{R}w_2$.

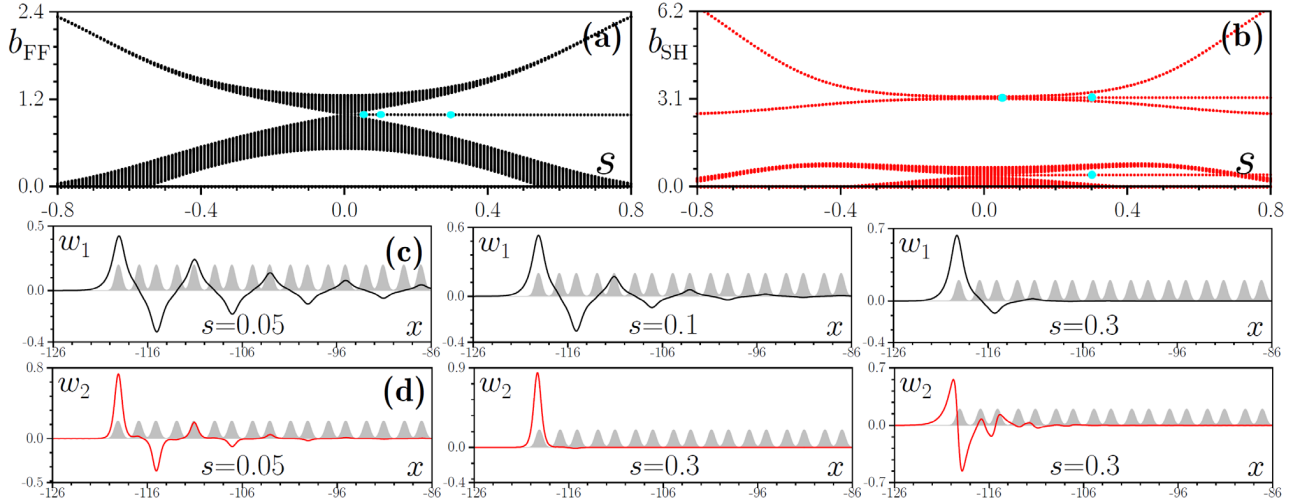

Fig. 1. Propagation constants of linear eigenmodes of the array vs waveguide shift $s$ for FF wave (a) and for SH wave (b). The spectrum for SH wave is obtained at $\beta=0$. Examples of profiles of linear topological FF (c) and SH (d) eigenmodes for different waveguide shifts indicated on the panels and corresponding to the cyan dots in panels (a) and (b), respectively. The profile of the array $\mathcal{R}(x)$ is shown schematically by gray shading. Here and below $a=2$, $d=0.5$, $p=2.5$.

Topological phase is introduced in SSH array by opposite shifts of the waveguides in each dimer that changes intra-cell and inter-cell coupling. Transformation of spectra with $s$ is illustrated in Fig. 1(a) for FF wave, and in Fig. 1(b) for SH wave at $\beta=0$. For $s\leq 0$ in nontopological regime all eigenmodes are extended, but for $s>0$, when inter-cell coupling becomes stronger than intra-cell one due to waveguide shifts, localized edge states appear in the topological gap. They show growing localization with increase of $s$ [see Fig. 1(c) for FF wave]. Edge states for SH wave are more localized for the same value of $s$ [Fig. 1(d)]. For $\beta=0$ one can see second topological gap in the SH wave spectrum, where edge states possess different symmetry inside waveguides [right column of Fig. 1(d)]. Since $\beta$ is included into second of Eqs. (2), the spectrum from Fig. 2(b) simply shifts downwards/upwards with increase/decrease of $\beta$.

Further we address edge solitons bifurcating from linear edge states under the action of $\chi^{(2)}$ nonlinearity. They can be found in the form $\psi_1=w_1e^{ibz}$, $\psi_2=w_2e^{2ibz}$, where propagation constant $2b$ of the SH wave is now defined by the propagation constant $b$ of the FF wave. We are interested only in solitons with $b$ belonging to the *topological gap* in the FF wave spectrum $b_{FF}$. The structure of the SH component is determined by the location of its propagation constant, $2b$, inside corresponding linear spectrum $b_{SH}-\beta$ for SH wave [here, for convenience, we use as $b_{SH}$ the spectrum calculated at $\beta=0$, Fig. 1(b)]. In Fig. 2 we superimpose on one $b(\beta)$ plot the allowed bands (gray areas) for FF wave that do not depend on $\beta$, and allowed bands (areas between red lines) for SH wave, whose borders can be defined using the condition $2b=b_{SH}-\beta$, where we take $b_{SH}$ calculated at $\beta=0$: these borders vary linearly with $\beta$. This figure is obtained for representative value of the shift $s=0.3$, when the array is in topological phase. One can obtain localized solitons only if $b$ from FF *topological gap* is taken such that SH wave simultaneously falls into its own forbidden *topological* or *nontopological gap*. Thus, at $\beta=0$ the SH wave can only belong to nontopological gap below band 2. In the interval $0.45 \leq \beta \leq 1.97$ the propagation constant of the SH wave may belong to topological SH gap between bands 1 and 2 meaning that both soliton components may have topological nature. At $\beta<-0.71$ the exotic situation is possible when SH component may fall into second topological SH gap.

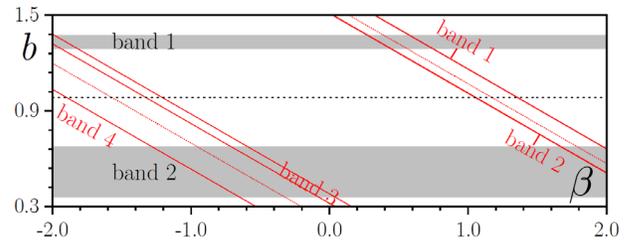

Fig. 2. Superimposed bands and gaps for FF and SH waves on the $(\beta,b)$ plane for $s=0.3$. Topological edge states in the gaps are indicated by the dotted lines. Bands for FF wave are shown gray, bands for SH wave are located between solid red lines.

To confirm these conclusions, we obtained exact soliton families from Eq. (1) using Newton method. To characterize them, we calculated total power $U=U_1+U_2=\int_{-\infty}^{\infty}(|\psi_1|^2+|\psi_2|^2)dx$ and power sharing $S_{1,2}=U_{1,2}/U$ between FF and SH components vs propagation constant $b$. Stability was analyzed by substituting perturbed solitons $\psi_m=(w_m+u_me^{\lambda z}+v_m^*e^{\lambda^* z})e^{imbz}$ $(m=1,2)$ into Eq. (1), linearizing it and solving corresponding linear eigenvalue problem to obtain perturbation growth rates $\lambda$ and profiles $u_m, v_m \ll w_m$.

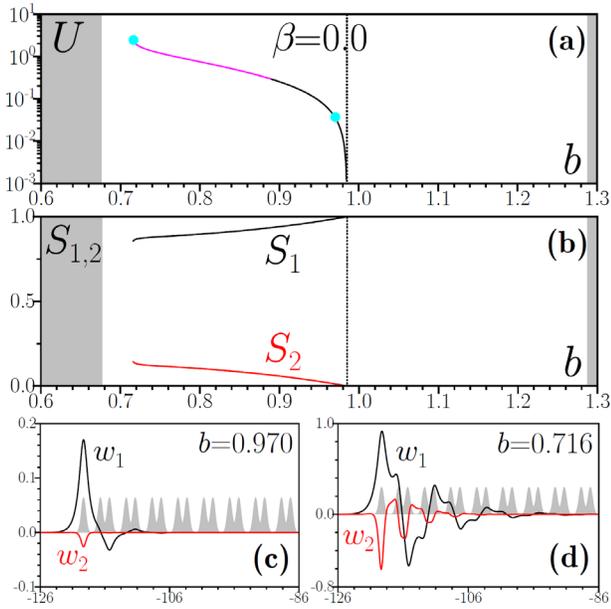

linear FF edge state (dotted gray line) towards decreasing $b$ values [Fig. 3(a)]. This *first type* of edge soliton is thresholdless, SH component in this state vanishes in bifurcation point, but becomes more pronounced when $b$ approaches lower edge of the FF gap [Fig. 3(b)]. When bifurcation occurs in the direction of decreasing $b$ values, as in this case, the SH component in the outermost guide is always negative. For selected $s$ the soliton is well localized near bifurcation point [Fig. 3(c)], but extends into array near lower FF gap edge [Fig. 3(d)]. FF soliton component inherits the structure of tails from linear edge state (staggered tails with amplitude changing its sign on each second waveguide). Considerable part of this soliton family is stable [black segments of $U(b)$ curve], while instabilities appear near the bottom of the gap (magenta segments).

More interesting situation occurs at larger values of $\beta \sim 0.9$, when SH bands in Fig. 2 "enter" into topological FF gap (i.e., propagation constant $2b$ of the SH wave can fall into SH topological gap). In this case, in addition to soliton bifurcating from FF edge state, one observes *second type* of edge soliton emanating from topological state in SH wave [see branch in Fig. 4(a) emerging from dashed red line associated with topological edge state in SH gap]. In such edge solitons of the second type FF wave vanishes, but SH wave remains nonzero in the bifurcation point [Fig. 4(c)]. Remarkably, now *both* FF and SH components feature structure of tails representative for topological edge states, see Fig. 4(e). Solitons of the first and second types can coexist and can be stable. Figure 4(a) shows also that the continuation of the family of the second type is found above upper red band (above band 1 for SH wave). The structure of tails of SH wave in this branch changes. This family may be stable even close to the FF gap edge, where soliton strongly extends into array [Fig. 4(f)].

Fig. 3. Total power $U$ (a) and power sharing $S_{1,2}$ between components (b) vs $b$ for topological edge solitons at $\beta=0$, $s=0.3$. Stable branches in (a) are shown black, unstable – with magenta color. Gray areas show allowed bands for FF wave. (c), (d) Edge solitons corresponding to cyan dots in (a).

The simplest situation for selected shift $s$ (for other $s>0$ values the picture is similar) is realized at $-0.71<\beta<0.45$, when topological FF gap in Fig. 2 is not "crossed" by SH bands. In this case, there exists the only family of topological solitons bifurcating from

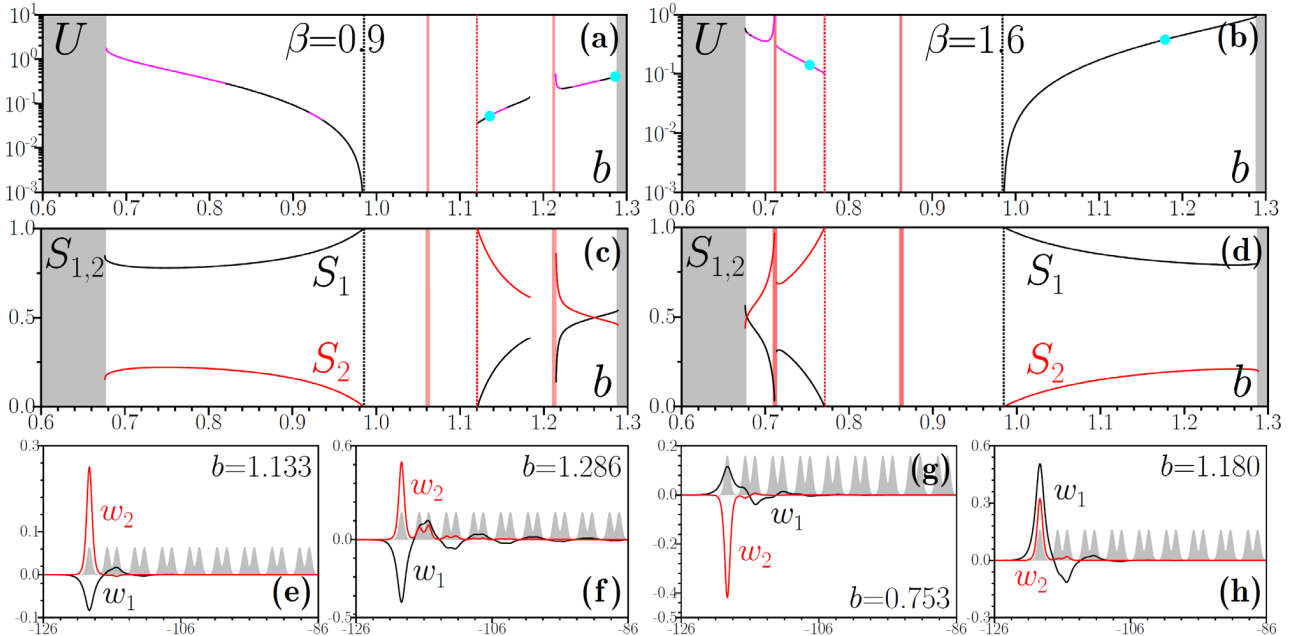

Fig. 4. Total power $U$ (a),(b) and power sharing $S_{1,2}$ (c),(d) versus $b$ for edge solitons at $\beta=0.9$ (a),(c) and $\beta=1.6$ (b),(d). Gray and pink areas indicate allowed bands for FF and SH waves, respectively, while vertical dotted lines indicate topological states in FF and SH waves. Stable branches in (a),(b) are shown black, unstable – with magenta color. Edge solitons corresponding to the cyan dots in $U(b)$ curves at $\beta=0.9$ (e),(f) and $\beta=1.6$ (g),(h). In all cases $s=0.3$.

Bifurcation picture changes qualitatively for $\beta>1.2$, when topological SH band "crosses" the level corresponding to FF edge state (see Fig. 2). In this case, thresholdless edge soliton of the first type emanating from FF edge state is found in the upper part of the topological gap, i.e. now it bifurcates in the direction of increasing $b$ values [see Fig. 4(b) and 4(d)] with the entire family being stable. SH component in soliton of the first type is now positive in the edge waveguide. In contrast, soliton of the second type bifurcates from SH edge state in the direction of decreasing $b$ values. This family features power threshold and is unstable in the topological SH gap, and only its continuation below band 2 for SH wave can have stable segments. We did not find solitons *between* black and red dotted lines associated with linear FF and SH topological edge states.

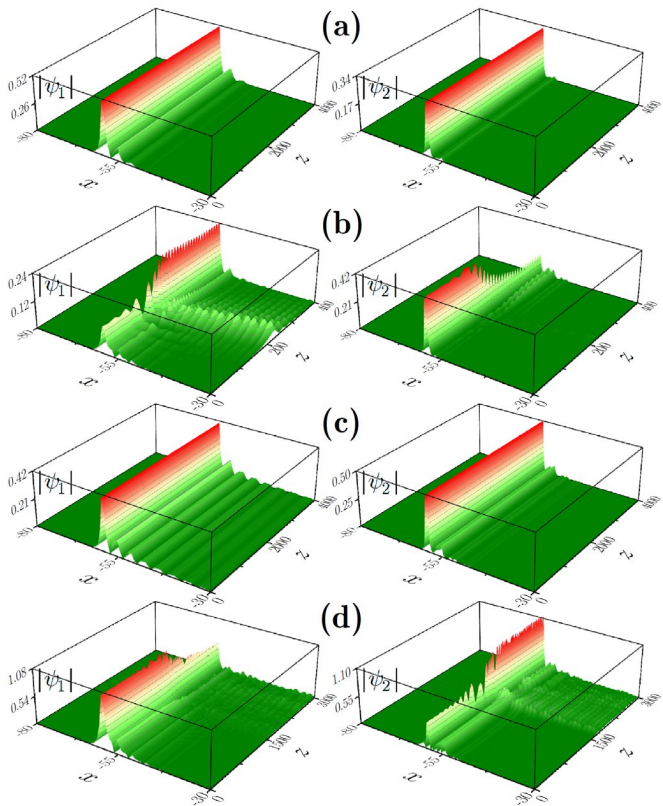

Fig. 5. Propagation dynamics of stable (a),(c) and unstable (b),(d) edge solitons at $b=1.180$, $\beta=1.6$ (a), $b=0.753$, $\beta=1.6$ (b), $b=1.286$, $\beta=0.9$ (c), and $b=0.745$, $\beta=-1.0$ (d). In all cases $s=0.3$.

The examples of stable propagation of well-localized and wide perturbed topological edge solitons are presented in Fig. 5(a), (c). Unstable representatives of topological soliton family, like soliton of the second type from Fig. 5(b) usually decay via progressively increasing oscillations of the FF and SH components. In Fig. 5(d) we also illustrate propagation of exotic topological soliton at $\beta=-1$, whose SH falls into second topological gap for SH wave – such states are usually unstable for our parameters.

**Funding:** Russian Science Foundation (grant 21-12-00096).
**Disclosures:** The author declares no conflicts of interest.
**Data availability.** Data underlying the results presented in this paper are not publicly available at this time but may be obtained from the authors upon reasonable request.